\documentclass[
 reprint, superscriptaddress, floatfix,
 amsmath,amssymb,
 aps,
 pra,
]{revtex4-2}

\usepackage{xcolor}
\usepackage{graphicx}
\usepackage{dcolumn}
\usepackage{bm}
\usepackage[colorlinks=true, allcolors=blue]{hyperref}
\DeclareUnicodeCharacter{2212}{-}
\begin{document}

\preprint{APS/123-QED}

\title{Hyperfine structure and isotope shifts of the \texorpdfstring{$^1P_1 \leftarrow{} ^{1}S_0$}{} transition in atomic zinc}

\author{David Röser}\thanks{These authors contributed equally to this work.}
\affiliation{University of Bonn, Nussallee 12, 53115 Bonn, Germany}
\author{J. Eduardo Padilla-Castillo}\thanks{These authors contributed equally to this work.}
\affiliation{Fritz-Haber-Institut der Max-Planck-Gesellschaft, Faradayweg 4-6, 14195 Berlin, Germany}
\author{Ben Ohayon}
\affiliation{The Helen Diller Quantum Center, Department of Physics, Technion-Israel Institute of Technology, Haifa, 3200003, Israel}
\author{Russell Thomas}
\affiliation{Fritz-Haber-Institut der Max-Planck-Gesellschaft, Faradayweg 4-6, 14195 Berlin, Germany}
\author{Stefan Truppe}
\affiliation{Fritz-Haber-Institut der Max-Planck-Gesellschaft, Faradayweg 4-6, 14195 Berlin, Germany}
\affiliation{Centre for Cold Matter, Blackett Laboratory, Imperial College London, London SW7 2AZ}
\author{Gerard Meijer}
\affiliation{Fritz-Haber-Institut der Max-Planck-Gesellschaft, Faradayweg 4-6, 14195 Berlin, Germany}
\author{Simon Stellmer}
\affiliation{University of Bonn, Nussallee 12, 53115 Bonn, Germany}
\author{Sid C. Wright}
\email[]{sidwright@fhi-berlin.mpg.de}
\affiliation{Fritz-Haber-Institut der Max-Planck-Gesellschaft, Faradayweg 4-6, 14195 Berlin, Germany}

\date{\today}
\begin{abstract}
We report absolute frequency, isotope shift, radiative lifetime and hyperfine structure measurements of the $^1P_1 \leftarrow$ 
 $^{1}S_0$ (213.8~nm) transition in Zn I using a cryogenic buffer gas beam. Laser-induced fluorescence is collected with two orthogonally oriented detectors to take advantage of differences in the emission pattern of the isotopes. This enables clear distinction between isotopes whose resonances are otherwise unresolved, and a measurement of the fermion hyperfine structure parameters, $A(^{67}$Zn)$=20(2)$~MHz and $B(^{67}$Zn)$=10(5)$~MHz. We reference our frequency measurements to an ultralow expansion cavity and achieve an uncertainty at the level of 1~MHz, about 1 percent of the natural linewidth of the transition.

\end{abstract}

\maketitle

\section{\label{sec:level1}Introduction}

The alkaline-earth-metal (AEM) elements are identified by two valence electrons and a $J=0$ electronic ground state. These two features give rise to a number of unique properties. Firstly, the level structure decomposes into singlet and triplet states, with broad transitions within each system and narrow intercombination lines between them. Just as in the helium atom, the lowest triplet states are metastable. Second, states with zero electronic spin are free of hyperfine structure. In addition, bosonic isotopes have even proton and even neutron numbers, leading to zero nuclear spin and absence of hyperfine structure in all electronic states. These properties enable a wealth of applications, including optical clocks \cite{Ludlow2015}, precision metrology \cite{Tarallo2014}, quantum computing \cite{Daley2008,Gorshkov2009,Saffman2020}, and Rydberg physics \cite{Simien2004,Browaeys2020}. In recent years, AEM elements have played a major role in the search for yet undiscovered scalar gauge bosons through high-precision isotope shift spectroscopy \cite{Berengut2018}, and various studies with neutral AEM atoms have been presented on this topic \cite{Miyake2019,Witkowski2019,Figueroa2022,Ono2022,Hofsaess2023,Ohayon2022}.

Alongside the AEM elements, and sharing these attractive properties, are the so-called Group-IIB elements zinc, cadmium, and mercury. The broad singlet ${}^1P_1 \leftarrow {}^1S_0$ transitions for these elements lie deep in the ultraviolet range of the spectrum, with natural linewidths in excess of 100~MHz. They possess a multitude of bosonic and fermionic isotopes, with the latter showing hyperfine structure. The resonance lines of the different isotopes are convoluted and often cannot be resolved in conventional Doppler-free spectroscopy. While Cd and Hg have already been employed for the development of optical clocks \cite{Yamaguchi2019, DeSarlo2016, Tyumenev2016}, there is very modest work towards this application with zinc thus far \cite{Bueki2021}, mainly limited by the available laser technology.
The wider chain of radioactive zinc isotopes is of interest for nuclear structure studies. For this reason, their isotope shifts have been measured in the triplet manifold~\cite{Wraith2017, Xie2019}, with new experiments ongoing~\cite{2020-ISOL, 2023-Coll}.

Here, we present high-resolution spectroscopy of the ${}^1P_1 \leftarrow{} {}^1S_0 $ transition near 213.8~nm in neutral zinc. Experiments were conducted over a two week campaign in which an ultralow expansion cavity and required deep ultraviolet optics (University of Bonn) were transported to an atomic beam machine at the Fritz Haber Institute in Berlin. Our measurements are based on laser-induced fluorescence of a cryogenic beam of atoms extracted from a helium buffer gas cell. We employ a two-detector method to clearly separate the contributions from the spin-zero bosonic isotopes from the spin-5/2 fermionic isotope, enabling use of natural abundance Zn. Isotope shifts and hyperfine interaction constants are determined with an uncertainty of order 1~MHz. This method provides a blueprint for measurements of hyperfine structure in strong optical transitions, and a convenient and direct way to measure the true collection solid angle of a fluorescence detector. Our approach can be readily adapted to other species with several naturally occurring isotopes, e.g., Sn, Ni, and applied in the study of radioactive nuclei.

\section{Experimental Setup}

\begin{figure*}
    \centering
    \includegraphics[width = 1.0\textwidth]{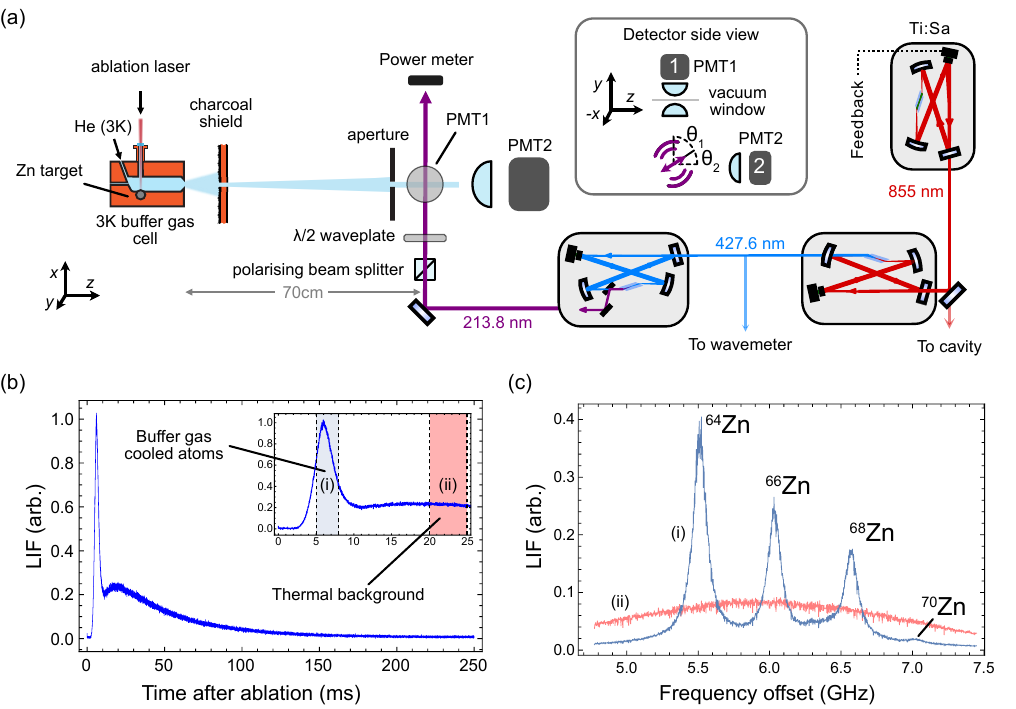}
    \caption{(a) Experimental setup for $^1P_1 \leftarrow{} ^1S_0$ laser induced fluorescence spectroscopy of Zn, showing the cryogenic buffer gas beam, and 213.8~nm laser system. We also show a side view of the detector geometry, showing the two photomultipliers used, and the angles $\theta_1$ and $\theta_2$ relative to linear polarisation angle of the excitation light. These angles determine the emission pattern observed at the two detectors. (b) A typical time-of-flight fluorescence trace observed at the $^{64}$Zn resonance. The inset shows a zoom-in of the region $0<t<25$~ms. Observation windows for the buffer gas cooled and thermal background components of the signal are shown by the shaded bars. (c) Fluorescence spectra for the observation windows in (b).}
    \label{fig:Setup}
\end{figure*}

Figure \ref{fig:Setup}(a) illustrates our experimental apparatus and laser system. We use a cryogenic buffer gas source to produce a cold, slow atomic beam of zinc. The atoms are produced by laser ablation of a solid Zn target (natural abundance), are cooled in the cell by collisions with a He gas at a temperature of 3~K, and exit the cell with a typical velocity of $140$~m/s along the $z$-axis. The ablation laser is fired at a rate of 1~Hz which sets the repetition rate for the experiments. At a distance 70~cm downstream of the cell exit, we excite the $^1P_1\leftarrow{} ^1S_0$ transition with a single probe laser beam near 213.8~nm, which intersects the atomic beam perpendicularly. A $2\times2$~mm square aperture restricts the range of transverse velocities in the atomic beam to below 1~m/s. 

Continuous wave laser light is produced by twice frequency doubling the infrared light of a Ti:Sa laser near 855.2~nm. Each frequency doubling stage consists of an enhancement resonator containing a nonlinear crystal; to reach 213.8~nm from 427.6~nm we use beta Barium Borate ($\beta$-BBO). The 855.2~nm light from the Ti:Sa laser is frequency stabilised either by referencing to a commercial wavemeter (High Finesse WS8-10 calibrated with a temperature-stabilised HeNe laser), or via an ultralow expansion (ULE) optical cavity (Menlo ORC) with a measured free spectral range 
 of 2.992184(30) GHz. We also record the intermediate 427.6~nm light on the wavemeter, since this light is immune from parasitic multi-mode content at the fundamental wavelength. The wavemeter option offers an absolute accuracy of about 30~MHz when measuring the 427.6~nm light, a resolution of about 1~MHz, and enables continuous scanning over the entire spectrum. Scanning via the reference cavity reduces the linewidth of the laser, and improves the linearity of the frequency axis. To do this, light at the fundamental wavelength of the Ti:Sa laser is coupled into a fiber phase modulator (EOM, Jenoptik) driven with two RF frequencies, $\nu_{PDH}=18$~MHz and $\nu_{scan} \sim$1~GHz. The phase-modulated light reflected from the cavity is collected on a fast photodiode, demodulated at the frequency $\nu_{PDH}$ 
which produces a Pound-Drever-Hall (PDH) signal with sharp zero crossing points when the laser frequency is at a cavity resonance $\nu_c$, or at $\nu_c \pm \nu_{scan}$. We lock the laser to the latter, and scan the laser frequency by varying $\nu_{scan}$. A camera is used to monitor the light transmitted through the cavity and ensure locking to the TEM$_{00}$ mode. This locking scheme enables continuous scanning of the Ti:Sa laser frequency up to one half of the cavity free spectral range, corresponding to 6~GHz at the 213.8~nm detection wavelength.

At the atomic beam machine, we purify the laser polarisation with a polarising beam cube, and control its linear polarisation angle relative to the direction of the atomic beam with a $\lambda/2$ plate. The probe light propagates along the $x$-axis and has a peak intensity $I=10^{-3} I_{sat}$, where $I_{sat}= \pi h c \Gamma/(3\lambda^3) = 1.5~$W/cm$^2$ is the two-level saturation intensity of the transition. We estimate that an atom travelling through the maximum intensity of the excitation light scatters five photons at resonance. The resulting laser-induced fluorescence (LIF) is collected and imaged onto two photomultiplier tubes (PMTs), whose photocurrents are delivered to separate transimpedance amplifiers and recorded as time-of-flight traces. The two PMTs are oriented to collect fluorescence emitted parallel and perpendicular to the direction of the atomic beam, as shown in figure \ref{fig:Setup}(a). The angle $\theta_i$ between the laser polarisation and the direction of detector $i$, illustrated in the inset to the figure, determines the portion of the fluorescence emission pattern collected by the two detectors. This enables discriminating between fermionic and bosonic isotopes \cite{Rasmussen1974,Baird1979,Hofsaess2023}. We record the laser power after the machine with a calibrated optical power meter and compensate for drifts in the probe intensity over a scan (typically $5-10$\%).

Figure \ref{fig:Setup}(b) shows a typical time-of-flight trace observed in detector 1 when exciting the $^{64}$Zn resonance. The signal comprises an initial intense peak from the buffer gas cooled atomic beam at roughly 5~ms, followed by an extended tail which appears for several tens of ms later. The extended tail consists of thermalised Zn atoms which leave the cell and collide with the vacuum walls; it persists even when direct line of sight from the source to the detector is blocked, and leads to a broad background signal in the fluorescence spectra, whose Doppler width is consistent with the laboratory temperature. Example spectra showing the two signal components are shown in Fig.~\ref{fig:Setup}(c). 

\section{Analysis of spectral lineshapes}

In the following, we discuss the lineshape models used for the fermionic and bosonic isotopes, which are important in fitting the experimental spectra. We hereon use $\nu_L$ to label the laser frequency and assume that the laser linewidth is much less than the natural linewidth of the transition.\\

\noindent \textit{Boson lineshape - } The bosonic isotopes, all with nuclear spin $I_N=0$, exhibit no hyperfine structure and the total angular momenta of the ground and excited states are $F=0$ and $F'=1$ respectively. The resonance line of a boson $b$ can be simply described with the line function,

\begin{equation}
S^{(b)} =\frac{\Gamma^2/4}{\Gamma^2/4+\Delta_b^2} [1-P_2(\cos\theta)g(\theta_C)] \hspace{0.3cm}.
\end{equation}

\noindent Here, $\Gamma/(2\pi)$ is the Lorentzian linewidth of the transition, $\Delta_b/(2\pi) = \nu_L - \nu_b$ is the detuning of the laser from the resonance frequency $\nu_b$, and $P_2(\cos\theta) = \frac{1}{2}(3\cos^2\theta-1)$ is the second Legendre polynomial, with $\theta$ the angle between the detection direction and the electric field of the linearly polarised excitation light. The factor $g(\theta_C) = \cos(\theta_C)\cos^2(\theta_C/2)$ corrects for the effect of the finite solid angle of the detection optics, with $\theta_C$ the half angle of a circular collection lens. For $\theta =0$, $\mathcal{S}^{(b)}\rightarrow 0$ as $\theta_C \rightarrow 0$, as would be expected from the well-known Hertzian dipole radiation pattern. Importantly, adjusting $\theta$ or $\theta_C$ changes the amplitude of the boson signal observed at the detector.\\

\begin{figure}
    \centering \includegraphics[width = \columnwidth]{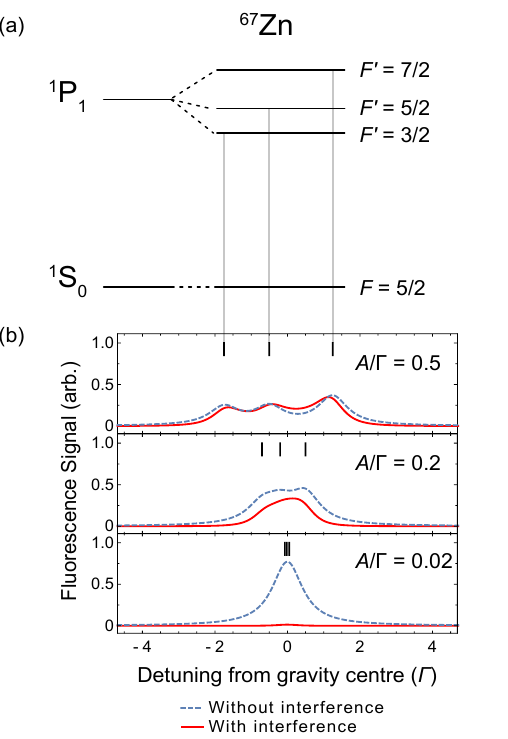}
    \caption{Simulated fluorescence spectra for $^{67}$Zn. (a) Level scheme labelling the total angular momenta $F,$ $F'$ for the ground and excited states respectively. (b) Simulated spectra for detection along $\theta = 0$, for different values of the ratio $A/\Gamma$ and with $B=0$. We show the results with and without quantum interference included in the calculation. The sticks above the spectra correspond to the energies of the levels in (a). As $A/\Gamma$ approaches zero (negligible hyperfine interaction), the emission pattern including interference converges to that of an ideal Hertzian dipole, meaning the emission is zero along $\theta=0$.}
    \label{fig:FermionSim}
\end{figure}

\noindent \textit{Fermion lineshape - } There exists a single naturally abundant fermionic isotope of Zn with nucleon number 67 and a nuclear spin $I_N=5/2$. The nuclear spin couples with the electronic angular momentum $J$ to give total angular momentum $F$, resulting in a single $^1S_0, F=5/2$ hyperfine level and three $^1P_1, F'$ excited levels with $F'=3/2,5/2,7/2$. These energy levels are shown in figure~\ref{fig:FermionSim}(a). We assume the hyperfine energies $E(F)$ are given by

\begin{equation}
    E(F) = \frac{A}{2}C + B\frac{\frac{3}{4}C(C+1) - I_N(I_N+1)J(J+1)}{2I_N(2I_N-1)J(2J-1)} \hspace{0.1cm},
\end{equation}

\noindent with $C=F(F+1)-I_N(I_N+1)-J(J+1)$. Here, $A = A(^1P_1$) is the interaction between the electronic and nuclear angular momentum in the excited state, and $B=B(^1P_1)$ is the quadrupole interaction coefficient.  

Following Brown et al. \cite{Brown2013}, the fluorescence spectrum of $^{67}$Zn, $S^{(f)}$ can be separated into three terms:

\begin{equation}
\begin{split}
    S^{(f)} = &  \frac{\Gamma^2}{4}\Big(\mathcal{A} + [
\mathcal{B}+ \mathcal{C}]P_2(\cos\theta) g(\theta_C)\Big) \hspace{0.3cm}, \\
\mathcal{A} =& \frac{1}{9}(\frac{2}{\Gamma^2/4 + \Delta_{3/2}^2} +  \frac{3}{\Gamma^2/4 + \Delta_{5/2}^2} +  \frac{4}{\Gamma^2/4 + \Delta_{7/2}^2}) \hspace{0.2cm}, \\
\mathcal{B} =& -\frac{1}{225}\frac{1}{\Gamma^2/4 + \Delta_{3/2}^2} \\
            & - \frac{64}{525}\frac{1}{\Gamma^2/4 + \Delta_{5/2}^2}  - \frac{2}{21}\frac{1}{\Gamma^2/4 + \Delta_{7/2}^2}  \hspace{0.3cm}, \\  
\mathcal{C} =& \Big[ -\frac{8}{45}\frac{1}{(\Gamma/2 + i\Delta_{7/2})(\Gamma/2-i\Delta_{3/2})} \\
            &-\frac{6}{35}\frac{1}{(\Gamma/2 + i\Delta_{7/2})(\Gamma/2 - i \Delta_{5/2})} \\
            &-\frac{1}{25}\frac{1}{(\Gamma/2 + i\Delta_{5/2})(\Gamma/2 - i \Delta_{3/2})} \Big] + c.c.
\end{split}
\label{eq:FermionLineFunction}
\end{equation}

\noindent Here, $\Delta_{F'}/(2\pi) = \nu-\nu_{F'}$ is the detuning of the laser from the excited state with total angular momentum $F'$; We assumed all Zeeman sublevels in the $^1S_0$ state are equally populated in the source and neglected optical pumping during the interaction with the probe light.  

Important for the experiments is the fact that when the hyperfine structure is barely resolved, the emission pattern and hyperfine structure become strongly coupled. This is illustrated by figure \ref{fig:FermionSim}b, which shows simulated fluorescence spectra along $\theta = 0$ for different ratios of $A/\Gamma$. Each panel compares equation \eqref{eq:FermionLineFunction} with the result when interference is removed from the model, i.e. $\mathcal{C}$ is deliberately set to zero. The calculations show that as $A/\Gamma \rightarrow 0$, interference between scattering paths is destructive, leading to complete suppression of the fluorescence along this direction. There is an intuitive explanation for this effect: when the hyperfine interaction with the nucleus becomes negligible, the emission pattern must converge to that of the (spin-less) bosonic isotopes. Conversely, one can produce the reverse effect in the bosonic isotopes by deliberately applying a magnetic field along $\theta=90^\circ$. This is the so-called Hanle effect \cite{Hanle1923, Mitchell1961} and, whilst understood for about a century, is often overlooked. The behaviour illustrated in figure \ref{fig:FermionSim}b shows that interference in the emission pattern of barely resolved lines contains useful information which can be used to constrain the hyperfine structure. The central spectrum in the figure, where $A/\Gamma = 0.2$, is near the value observed in the experiments, and results in a total span of the $^1P_1$ levels of 1.2~$\Gamma$. This should largely prevent optical pumping to the $m_F = \pm 5/2$ states when driving the $F'=3/2\leftarrow F=5/2$ transition, which would be significant in the case of well-resolved lines.\\

\noindent \textit{Combined line function - } The total fit function used in this study is given by,

\begin{equation}
\begin{split}
    S^{(tot)} = & a_{67}S^{(f)} + \sum _b a_b S^{(b)}  \\
                & + a_{bg}e^{-(\nu_L-\nu_{bg})^2/(2 w_{bg}^2)} _{} \hspace{0.3cm}.
    \label{eqn:FullFitFunction}
\end{split}
\end{equation}

\noindent Here, $a_{67}$ and $a_{b}$ represent the relative abundance of the fermionic and bosonic isotopes respectively. The final term in equation \eqref{eqn:FullFitFunction} approximates the residual thermal background in the spectrum, whose amplitude $a_{bg}$ is typically 5 to 10 percent of the $^{64}$Zn resonance peak. The centre frequency $\nu_{bg}$ and width parameter $w_{bg}$ can be either fitted as free parameters, or introduced as fixed parameters by first fitting the data at late arrival times when only the thermal background component is present. The fitted values for the isotope shifts in these two cases are consistent within the statistical error of the fits.

\section{Results}

\subsection{Determination of the fermion hyperfine structure by a two-detector method}
\label{sec:wavemeter}

\begin{figure*}
    \centering
    \includegraphics[width =2\columnwidth]{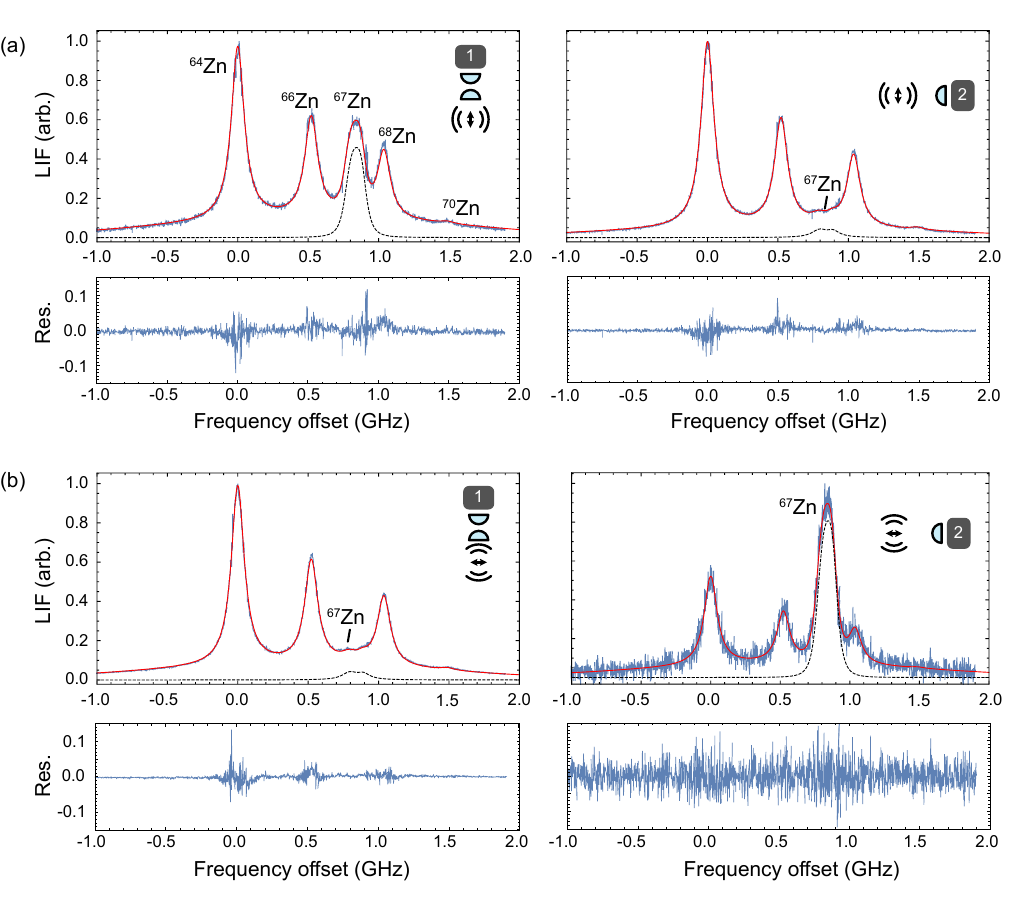}
    \caption{Polarisation sensitive fluorescence detection of Zn isotopes. Each spectrum is labelled by the probe laser polarisation and detector configuration. The relative intensities of the boson and fermionic resonances are strong functions of the angle $\theta$ between the laser polarisation and the detector direction, and the solid angle of the collection optics. Blue lines show experimental data, red solid lines are fits as described in the text, and black dashed lines show the fitted fermion lineshape. Underneath each spectrum the residuals are shown in a separate plot (`Res'). (a) Laser polarisation along the $y$-axis. (b) Laser polarisation along the $z$-axis.}
    \label{fig:wavemeterData}
\end{figure*}

Figures \ref{fig:wavemeterData}(a) and (b) show two sets of spectra obtained using the High Finesse wavemeter as a frequency reference. The data constitutes two separate scans where the input polarisation of the laser is along the $y$-axis (panel (a)) and along the $z$-axis (panel (b)), and for each panel we show the fluorescence spectrum recorded by the two detectors $1$ and $2$. For clarity, each is labelled with a schematic showing the laser polarisation, the detector orientation and the dominant emission pattern for the bosonic isotopes. The different emission pattern of the fermionic $^{67}$Zn isotope (relative natural abundance $4.1\%$) dramatically increases its visibility in detector $i$ when $\theta_i = 0$. We show the fitted fermion lineshape with a black dashed line in each panel to illustrate this effect.

We use these spectra to determine solid angles of the collection optics and hyperfine structure of the $^{67}$Zn isotopes. The four spectra in figure \ref{fig:wavemeterData} were fitted as a single dataset, fixing the detection angles $\theta_i$ to their values in the experiment, and enforcing the natural abundance of Zn isotopes \cite{Rosman1998}. This fixes the relative peak heights in each spectrum so that the detector solid angles $\theta_{C,1}, \theta_{C,2}$, and the hyperfine structure constants $A$ and $B$ of the fermionic isotope, can be determined. All resonance frequencies $\nu_b$ for all bosons $b$, $\nu_{1/2,3/2,5/2}$ for the fermionic $^{67}$Zn isotope, and a common Lorentzian linewidth $\Gamma$ are shared fit parameters between the datasets. From this data, we conclude $\theta_{C,1} = 0.281\pm0.005$, $\theta_{C,2}= 0.145\pm0.005$ radians. The uncertainties are the range of values obtained when fitting the data with various reasonable assumptions, such as fixing the values of $w_{bg}$ and $\nu_{bg}$ in the fit function using the signal at late arrival times. The value of $\theta_{C,2}$ is very close to the half angle subtended by the collection lens at the fluorescence region, $0.156(5)$~radians. The value of $\theta_{C,1}$ is significantly below the half angle subtended by its in-vacuum collection lens, $0.43(1)$~radians, and consistent with this lens being placed about $5$~mm too close to the atomic beam, a result of incorrectly extrapolating the focal length from the visible to the deep ultraviolet. 

For the hyperfine interaction parameters, we obtain  $A(^{67}$Zn$) = 20\pm 2$~MHz and $B(^{67}$Zn$)=10\pm 5$ MHz. Fitting the data with $B=0$ returns $A=21$~MHz but noticeably reduces the goodness of the fit near the $^{67}$Zn peak. Fitting to a model which ignores interference may be done by simply setting $\mathcal{C} = 0$ in equation \eqref{eq:FermionLineFunction}; this gave the best fit values $A = 9.5\pm1.8$~MHz, $B=0.6\pm2.2$~MHz, $\theta_{C,1} = 0.37$ and $\theta_{C,2} = 0.20$, and line centres consistent with the full interference model. In this case the fitted value of $\theta_{C,2}$ is unphysically large, and the fit residuals clearly indicate that only the interference model can adequately describe the signal observed in both detectors. 

\subsection{High resolution measurements with the cavity}

\begin{figure}
    \centering
    \includegraphics[width = \columnwidth]{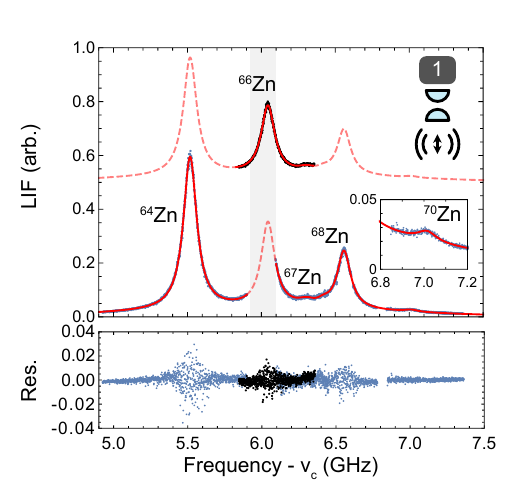}
    \caption{Fluorescence spectra taken with the probe laser locked to the ULE cavity. The shaded box shows the region near the midpoint of the cavity resonances in which the lock fails. In the upper spectrum, shown with a deliberate offset on the $y$-axis, the light delivered to the cavity was first frequency-shifted by 90~MHz. This enables a continuous scan over the $^{66}$Zn resonance. The two spectra are fitted as a single dataset to recover the isotope shifts as discussed in the text. Dashed lines show the fit functions where data is not present for each spectrum. The plot underneath shows the residual from the fit function for the two datasets.}
    \label{fig:cavityData}
\end{figure}

Having constrained the fermion hyperfine structure and solid angle of the collection optics, we proceeded to scan the laser via the ULE reference cavity to more accurately determine the isotope shifts. Figure~\ref{fig:cavityData} shows spectra obtained when scanning $\nu_{scan}$ with the laser locked to the ULE cavity, and with the probe laser horizontally polarised. The scan rate corresponds to approximately 0.8~MHz/s for the 213.8~nm probe light, where the frequency $\nu_{scan}$ was measured near the time of the ablation laser pulse, and then stepped discretely after each measurement. By happenstance, the $^{66}$Zn resonance appeared almost exactly at the midpoint between two cavity resonances, where the locking method fails. We therefore frequency-shifted the Ti:Sa laser light by $90$~MHz before delivery to the cavity with an acoustic optic modulator, moving the unstable lock point by 360~MHz in the deep ultraviolet. The upper (lower) dataset in the figure is taken with (without) the frequency shifting method applied, on the same day but ablating different spots on the Zn target, which enabled measuring all isotopes. We fit the two spectra as a single dataset with shared resonance line positions in a Monte Carlo routine, where the values of $\theta_{C,1},\theta_{C,2}, A$ and $B$ are drawn from uniform distributions whose ranges are given by the limits constrained in section~\ref{sec:wavemeter}. Enforcing the relative natural abundance of Zn in the fits leads to small changes in the best fit values compared to allowing the line intensities to float, and we include this when estimating the uncertainties. We combine the best-fit values and errors for the isotope shifts to give a weighted mean and statistical error for these parameters, and assume a $1$~MHz systematic frequency uncertainty which derives from the $\sim 200$~kHz uncertainty of $\nu_{scan}$ and the frequency shifting AOM, considering the two successive stages of frequency doubling. Doppler shifts due to slight misalignment of the probe laser light contribute an uncertainty $\sim 2$~MHz to the absolute resonance frequencies, and negligibly to the isotope shifts. Recoil from absorption of the probe laser light leads to a Doppler shift of roughly 0.6~MHz across the detection volume, and the differential shift across the range of isotopes is an order of magnitude smaller. We neglect this contribution to the isotope shift uncertainty. The ambient magnetic field in the detector was measured as below 0.3~Gauss, corresponding to an upper bound to the line shape broadening of 0.5~MHz.

\section{Discussion}

Table \ref{tab:Results} summarises the results of our measurements, and compares them to the available literature values. Our final values for the isotope shifts are presented relative to the $^{64}$Zn resonance, since this isotope is of highest abundance and its line centre has the smallest statistical uncertainty. We combine measurements from the cavity and spectra taken using the wavemeter, and include a systematic frequency error of $2.1$~MHz for the wavemeter values, which derives from directly comparing frequency intervals measured by the cavity scan method with the wavemeter. Our values are two orders of magnitude more precise than previous measurements, also given in the table. Our value of $A(^{67}$Zn$)$ is consistent with the value measured by Kowalski and Tr\"{a}ger~\cite{Kowalski1976}, by level crossing spectroscopy of enriched $^{67}$Zn. However, this study was unable to experimentally constrain the value of $B(^{67}$Zn$)$ and we therefore recommend the values from our measurements.

\begin{table*}[tbp]\label{tab:ISs}
\caption{
Summary of the results obtained for the $^1P_1 \leftarrow{} ^1S_0$ transition in Zn. Hyperfine constants and the radiative lifetime refer to the excited state. Results are given in MHz unless otherwise stated. CG = centre of gravity.
} 
\begin{ruledtabular}
\begin{tabular}{l c r r r r r r}

 & \multicolumn{1}{c}{This work} &
\multicolumn{1}{c}{Ref.\cite{1950-CGK}}  &
\multicolumn{1}{c}{Ref.\cite{1958-HL}} &\multicolumn{1}{c}{Ref.\cite{Kowalski1976}} &\multicolumn{1}{c}{Ref.\cite{Gullberg2000}} &\multicolumn{1}{c}{Ref.\cite{Doidge1995}}\\ 
\hline
 $\nu_{66} - \nu_{64}$& 525.0(3.0) & $480(60)$ & $540(60)$\\
 $\nu_{67}^\mathrm{(CG)} - \nu_{64}$& 835(4) & - & - &&\\
 $\nu_{68} - \nu_{64}$ &  1039.8(1.7) & $989(60)$ & $960(85)$\\
 $\nu_{70} - \nu_{64}$& 1495(4) & -  & - &&\\
 $A(^{67}$Zn$)$ & 20(2) & & & $17.7(5)$ && \\
 $B(^{67}$Zn$)$& 10(5)  & & & - && \\
 $\nu(^1P_1-{}^1S_0)$/cm$^{-1}$& $46745.407(2)$ & & & & $46745.404(2)$&&\\
 $\tau$/ns & 1.440(18) & & & & & 1.40(3)\\
 \end{tabular}
\end{ruledtabular}
\label{tab:Results}
\end{table*}

The absolute frequency of the $^{64}$Zn resonance measured through our experiments is $1,401,391.66(6)$~GHz ($46745.394(2)$~cm$^{-1}$). The isotope-averaged line centre, $\nu(^1P_1-{}^1S_0)$, is computed as the average of the line centres weighted by isotopic abundance and given in the table. The uncertainty in our value is dominated by the wavemeter accuracy specified by the manufacturer. We find excellent agreement with the results of (isotope-unresolved) hollow cathode lamp measurements presented in Ref.~\cite{Gullberg2000}.

Our best fit value of the Lorentzian linewidth is $\Gamma/(2\pi) = 110.5(1.4)$~MHz, where the error derives from the standard deviation of the Monte Carlo fitted values combined with the frequency uncertainty from the cavity scanning method. Fitting with a Voigt lineshape did not change the value of $\Gamma$ within the uncertainty of the fit. The radiative lifetime $\tau=1/\Gamma$ given in the table is consistent with the weighted average of 5 measurements collated by Doidge~\cite{Doidge1995}, and is a factor of two more precise than previous measurements. 

\begin{figure}
    \centering
    \includegraphics[width = 1.0\columnwidth]{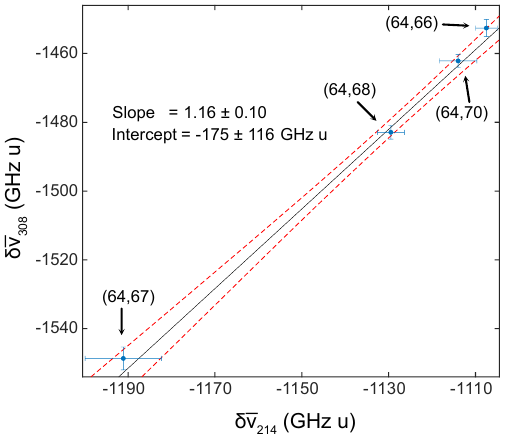}
    \caption{A King plot of the $^1P_1 \leftarrow{} ^1S_0$ (213.8~nm, this work) and $^3P_1 \leftarrow{} ^1S_0$ (308~nm, ref.~\cite{Campbell1997}) transitions in Zn. The black solid line shows a linear fit to the data as discussed in the text. Red dashed lines indicate the $68\%$ confidence interval of the fit.}
    \label{fig:KP}
\end{figure}

\textit{King plot} - We combine our isotope shift results with values reported for the $^3P_1 \leftarrow{} ^1S_0$ intercombination transition \cite{Campbell1997} on a King plot as follows. We calculate the reduced isotope shifts $\delta \bar{\nu}^{A,A'}=\delta \nu / \mu^{A,A'}$ with $\mu^{A,A'}$ the inverse nuclear mass difference between isotopes $A$ and $A'$, and present the data in figure \ref{fig:KP}. The data is fitted to a linear relationship according to the recipe described in Ref.~\cite{2023-Zn}. Briefly, we first define a mixing matrix to shuffle the 308~nm reduced isotope shifts so that they are referred to $^{64}$Zn. We then calculate the covariance matrices, taking into account the mixing matrix and the reported errors. The most-probable values of the adjusted parameters, the intercept and slope, are found by minimizing a generalized $\chi^2$ test statistic. To assign confidence intervals to the fitted parameters, we perform a Monte-Carlo estimation procedure of repeated measurements drawn from a normal distribution centred at the most probable fitted values. The most probable value and $68\%$ confidence interval of the fitted line is plotted in figure \ref{fig:KP}.

An additional hyperfine interaction between the different electronic states in Zn would additionally shift the gravity centre of the $^{67}$Zn resonance lines, $\nu_{67}^{(\mathrm{CG})}$, relative to the bosons, and is often referred to as an ``off-diagonal" hyperfine interaction. Such shifts should be detectable in a King plot as deviations of the fermionic isotopes from linear fits of the bosonic isotope data (see for example ref.~\cite{Miyake2019}). 
To test for this, we repeat the fitting procedure without the (64,67) pair, and calculate its predicted values in each iteration of the Monte-Carlo procedure to obtain its distribution. The resulting $68\%$ confidence interval for the frequency difference $\nu_{67}^{(\mathrm{CG})} - \nu_{64}$ is from 818 MHz to 834 MHz, which agrees with our measured value (835(4)~MHz) within its uncertainty. We infer that the difference in off-diagonal hyperfine shifts of the $^{67}$Zn resonances, for this pair of transitions, is less than $10\,$MHz. This closely follows results of isotope shift measurements for the lowest lying, and analogous, transitions in cadmium~\cite{Hofsaess2023}, another Group-IIB element. It may be of interest considering recently observed strong hyperfine mixing effects for higher-lying transitions in zinc in the CRIS experiment~\cite{2023-Coll}.

\section{Summary and outlook}

We have reported isotope shifts, radiative lifetime and hyperfine structure measurements for the $^1P_1 \leftarrow{} ^1S_0$ transition in neutral Zn by cw laser-induced fluorescence spectroscopy of an atomic beam.
Our measurements considerably improve upon the published literature for this transition, and contribute to the study of the $^{67}$Zn nucleus, where unexpected isotope shifts have recently been observed in collinear laser spectroscopy at the ISOLDE facility~\cite{2023-Coll}. 
With its multitude of spin-zero bosonic isotopes and various narrow optical transitions, zinc is a candidate for further isotope shift spectroscopy at the sub-kHz level.

The two-detector method and analysis procedure described here has several advantages. First, it enables reliably extracting hyperfine parameters from barely resolved peaks. Second, it enables tuning of the fermion contribution to spectra of mixed isotopes within a single measurement run. This approach allows disentangling otherwise overlapping lines and in the case of atomic Zn studied here, enables a measurement of the isotope shifts and hyperfine structure at the $\sim$1~MHz level. The approach can readily be adopted to other elements which feature broad transitions and many isotopes, and may be particularly beneficial in accelerator-based experiments, where measurements typically come with a large overhead.  

\begin{acknowledgments}
We thank Sebastian Kray and the mechanical workshop of the Fritz Haber Institute for expert technical assistance.
S.W. thanks Clara Bachorz for critical reading of the manuscript. 
We gratefully acknowledge financial support from the European Research Council (ERC) under the European Union’s Horizon 2020 Research and Innovation Programme (CoMoFun, Grant Agreement No.~949119, S.T.; ``quMercury", GA No.~757386, S.S.), from the European Commission through project 101080164 ``UVQuanT" (S.T. and S.S.), and from Deutsche Forschungsgemeinschaft (DFG) through grants 414709674 and 496941189 and through the Cluster of Excellence ML4Q (EXC 2004/1 - 390534769). 
B.O. is thankful for the support of the Council for Higher Education Program for Hiring Outstanding Faculty Members in Quantum Science and Technology. 
\end{acknowledgments}

\providecommand{\noopsort}[1]{}\providecommand{\singleletter}[1]{#1}%

\end{document}